\documentclass{astlb}

\usepackage{graphicx}
\usepackage{color}
\usepackage[hyperfootnotes=false,draft]{hyperref}

\definecolor{darkblue}{rgb}{0,0,0.9}

\def\ps1{\emph{Pan-STARRS1}}

\def\cl2305{SRGe\,CL2305.2$-$2248}

\begin{document}


\title{Mass estimation of the very massive galaxy cluster SRGe~CL2305.2-2248 from strong lensing}

\author{I.M.~Khamitov\email{irek\_khamitov@hotmail.com}\address{1,2},
  I.F.~Bikmaev\address{2,3,11},
  N.S.~Lyskova\address{4,8},
  A.A.~Kruglov\address{4},  
  R.A.~Burenin\address{4,11},
  M.R.~Gilfanov\address{4,5},
  A.A.~Grokhovskaya\address{6,7},
  S.N.~Dodonov\address{6,7,11},
  S.Yu.~Sazonov\address{4},
  A.A.~Starobinsky\address{9},
  R.A.~Sunyaev\address{4,5},
  I.I.~Khabibullin\address{10,4,5},
   E.M.~Churazov\address{4,5}
  \addresstext{1}{TÜBİTAK National Observatory, Antalya, Turkey}
  \addresstext{2}{Kazan Federal University, Kazan, Russia}
  \addresstext{3}{Academy of Sciences of The Republic of Tatarstan, Kazan, Russia}
  \addresstext{4}{Space Research Institute RAS (IKI), Moscow, Russia}
  \addresstext{5}{Max Planck Institute for Astrophysics, Garching, Germany}
  \addresstext{6}{Special Astrophysical Observatory of the Russian Academy of Sciences, Nizhnij Arkhyz, Russia}
  \addresstext{7}{Institute of Applied Astronomy RAS, St. Petersburg, Russia}
  \addresstext{8}{P.N.Lebedev Physical Institute RAS, Astro Space Center, Moscow, Russia}  
  \addresstext{9}{Landau Institute for Theoretical Physics RAS, Chernogolovka, Russia}
  \addresstext{10}{Observatory of the Ludwig Maximilian University of Munich, Munich, Germany}
  \addresstext{11}{Sternberg Astronomical Institute, Moscow State University, Moscow, Russia}
}

\shortauthor{Khamitov et al.}

\shorttitle{Mass estimation of \cl2305 cluster from the SRG/eROSITA survey}


\submitted{December 1, 2021}

\begin{abstract}
  The galaxy cluster \cl2305 (SPT-CL \, J2305 $ - $ 2248, ACT-CL \, J2305.1 $ - $ 2248) is one of the most massive clusters at high redshifts ($ z \simeq 0.76 $) and is of great interest for cosmology. For an optical identification of this cluster, deep images were obtained with the 1.5-m Russian-Turkish telescope RTT-150. Together with the open archival data of the Hubble Space Telescope, it became possible to identify candidates for gravitationally lensed images of distant blue galaxies in the form of arcs and arclets. The observed giant arc near the brightest cluster galaxies allowed us to estimate the radius of the Einstein ring, which is $ 9.8 \pm 1.3 $ arcseconds. 
  The photometric redshift of the lensed source was obtained ($ z_s = 2.44 \pm 0.07 $).  Its use in combination with the Einstein radius estimate made it possible to independently estimate the \cl2305 mass. It was done by extrapolating the strong lensing results to large radii and using the model density distribution profiles in relaxed clusters. 
  This extrapolation leads to mass estimates $ \sim 1.5-3 $ times smaller than those obtained from X-ray and microwave observations. A probable cause for this discrepancy may be the process of cluster merging, which is also confirmed by \cl2305 morphology in the optical range.

  
\end{abstract}

\section{Introduction}

The very massive cluster of galaxies \cl2305  was detected in X-rays from the results of the first all-sky survey (completed in June 2020) with the eROSITA telescope on board the SRG space observatory \citep{2021A&A...647A...1P, 2021A&A...656A.132S}.
Based on the results of optical observations with the Russian-Turkish 1.5-m telescope RTT-150 and the 6-m BTA telescope, an optical identification of the cluster was carried out, and a spectroscopic measurement of the redshift of the cluster was obtained, $ z = 0.7573 $ \citep{br21_pazh}. 
The presence of strong gravitational lensing in the
cluster field is noted in the paper by \cite{2020ApJS..247...25B} (see their Table~8) on the early detection of this
cluster (SPT-CL J2305-2248) in the millimeter band
of the South Pole Telescope survey. 
In this work, we carried out a photometric estimate of the redshift of the lensed source and obtained an independent estimate of the mass of the cluster within the Einstein ring. 
Assuming the Navarro-Frenk-White density distribution \citep{1996ApJ...462..563N}, we derived the cluster mass estimate at $ R_{500} $.
In our estimates we assume the standard cosmological model $ \Lambda $CDM with the following parameters: $ \Omega_m = 0.3 $, $ \Omega_{\Lambda} $ = 0.7, $ H_0 = 70 $ km/s/Mpc.

\section{Photometric observations}

Deep direct images of the \cl2305\ cluster field were obtained on the Russian-Turkish 1.5-m telescope RTT-150 as a part of the ground-based support of the X-ray survey with the eROSITA telescope onboard the Spectrum-Roentgen-Gamma (SRG) space observatory. The observations were performed in August 23-26, 2020, in the SDSS \emph{g,r,i,z} filters. An Andor iKon-L 936 BEX2-DD-9ZQ CCD camera 2Kx2K (2048x2048) pixels with a resolution element of 0.\arcsec326 was used as a detector. The quantum efficiency of the CCD is about 90\% and higher in the wavelength range from 4000\AA~to 8500\AA. Table~\ref{tab:log} shows a log of  observations made. The total moderate quality exposure times were 13200~s, 13800~s, 8400~s, and 8400~s in the SDSS filters \emph{g,r,i,z}, respectively. The total exposure in each filter was divided into 600~s exposures, between which the telescope pointing axis was shifted by 10-20\arcsec\ in an arbitrary direction. Processing of the direct images was performed in a standard way, using the  \emph{IRAF} software, as well as our own software, using a standard set of calibrations.

The photometric calibration of the images was obtained using observations of the photometric standards \citep{Smith02}.

\begin{table}
\caption{Observation log of the galaxy cluster \cl2305 performed at RTT-150. F - SDSS filter, N - number of images, T - total exposure in seconds, $\sigma$ - the seeing in arcseconds.}
  \label{tab:log}
  \vskip 2mm
  \renewcommand{\arraystretch}{1.05}
  \renewcommand{\tabcolsep}{0.35cm}
  \footnotesize
     \begin{tabular}{ccccc}
     \noalign{\vskip 3pt\hrule\vskip 5pt}
    Date	& 	F	&	N	&  T	  &	$\sigma$ \\
            &  (SDSS)     &   	&  (s)  &	($\arcsec$) \\
\hline
2020-08-23	&   g	&   4	&   2400  &  2.4 \\
2020-08-23	&   r	&   4	&   2400  &  2.0 \\
2020-08-23	&   i	&   4	&   2400  &  1.7 \\
2020-08-23	&   z	&   4	&   2400  &	 1.7 \\
2020-08-24	&   g	&   4	&   2400  &  1.6 \\
2020-08-24	&   r	&   4	&   2400  &  1.7 \\
2020-08-24	&   i	&   4	&   2400  &  1.6 \\
2020-08-24	&   z	&   5	&   3000  &  1.6 \\
2020-08-25	&   g	&   4	&   2400  &  2.0 \\
2020-08-25	&   r	&   5	&   3000  &  2.2 \\
2020-08-25	&   i	&   6	&   3600  &  2.1 \\
2020-08-25	&   z	&   5	&   3000  &  2.0 \\
2020-08-26	&   g	&   10	&   6000  &  2.0 \\
2020-08-26	&   r	&   10	&   6000  &  2.2 \\
\noalign{\vskip 3pt\hrule\vskip 3pt}
\end{tabular}
\end{table}

\label{sec:obs}

\section{Selection of  lensed images}

\begin{figure}
  \centering

  \includegraphics[width=\columnwidth]{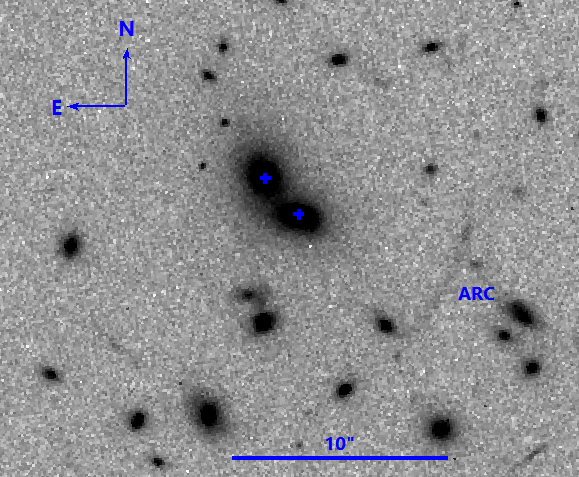}

  \hspace{1pt}
  \caption{HST-image of the cluster field. The giant arc is clearly visible. The positions of the centers of the two brightest cluster galaxies are marked with crosses.}
  \label{fig:HST}
  
\end{figure}

For the task of an  optical identification of the  cluster  under consideration,  deep images with a limiting magnitude as faint as 23-24$^{m}$ in the 4 bands \emph{g,r,i,z} of the SDSS photometric system  were obtained at the RTT-150 telescope. At redshift $ z = 0.76 $ the main radiation flux from the cluster  falls into the \emph{i} and \emph{z} bands. The images in the \emph{g,r} bands were used to cut off the background galaxies at smaller redshifts. The number of background galaxies separated by colors \emph{(g-r)} and  \emph{(r-i)} in the RTT-150 images and that did not belong to the cluster in the field of 9 x 9 arcminutes  was about 1000. An analysis of the images showed that, in addition to the large number of background galaxies detected in all \emph{g,r,i} bands  in regions close in position to two brightest cluster galaxies (BCG), about 10 blue sources, visible only in the \emph{g} filter, are detected. Given that \cite{2020ApJS..247...25B} noted the presence of strong gravitational lensing in the cluster field, we suggested that the blue sources in the RTT-150 frames near the BCG could be associated with the gravitational lensing features (arcs).
To select the lensed images of a distant source, we used deep RTT-150 images of the cluster field obtained in \emph{g,r,i,z} bands  and high spatial resolution images from the Hubble Space Telescope (HST) open data archive \citep{2016AJ...151..134W} obtained on April 23, 2018 in F110W filter with the WFC3/IR detector, an exposure time of 758.81 s (Fig.~\ref{fig:HST}), and in the F200LP filter with the WFC3/UVIS detector, an exposure time of 741 s.

\begin{figure}
  \centering

  \includegraphics[width=\columnwidth]{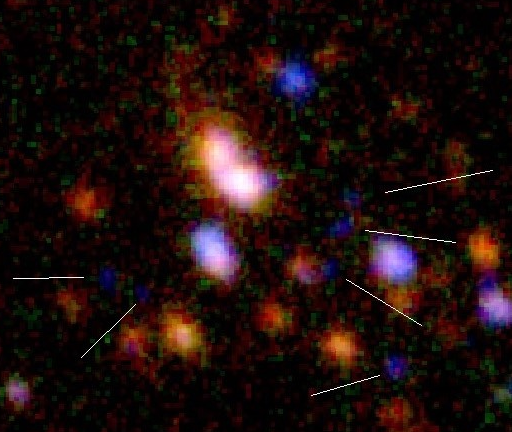}

  \hspace{1pt}
  \caption{The color image of the cluster field. The pseudo-colors are based on RTT-150 data (\emph{g} - blue, \emph{r} - green, \emph{i} - red). The selected blue sources in the form of  arcs and arclets in the HST image are marked.}
  \label{fig:RTT150_ARC}
  
\end{figure}

\begin{figure}
  \centering
   \includegraphics[width=\columnwidth]{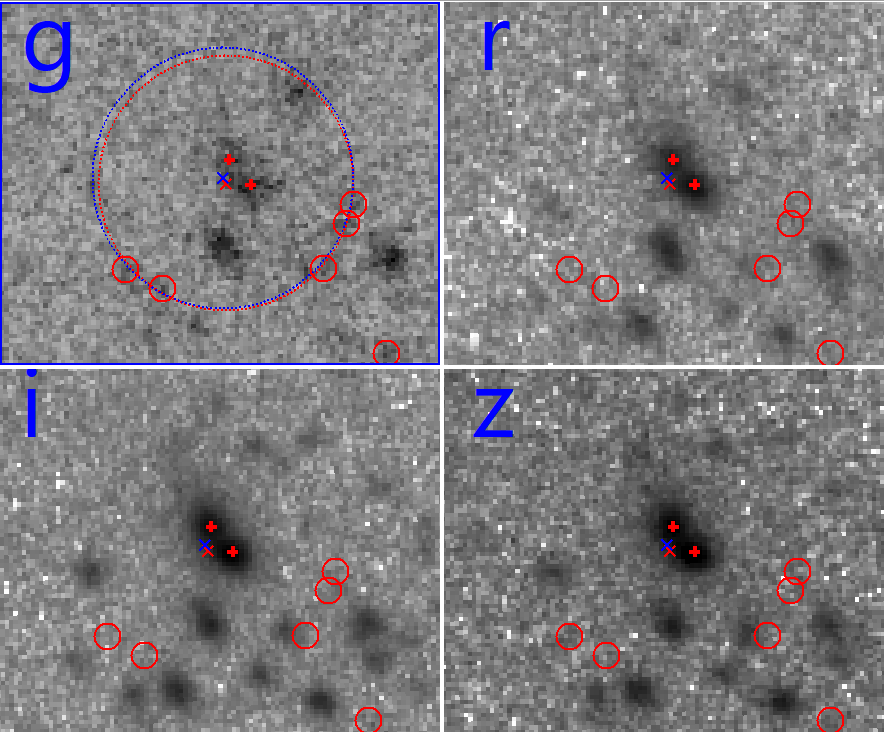}
  \caption{Deep $30\arcsec \times 30\arcsec$ \emph{g,r,i,z}-images of the field around \emph{BCG}. The positions of the two brightest galaxies of the cluster are marked with red crosses.}
  \label{fig:SL_griz}
\end{figure}

The lensed candidates were selected by the morphology of the sources in the HST image and by brightness excess  in the \emph{g} band compared to the \emph{r,i,z} bands. Fig.~\ref{fig:RTT150_ARC} shows the selected candidates in the form of arcs and arclets. The brightness of the sources is close to the detection limit in the \mbox{\emph{g}-image} and is  $\sim 24^{m}$.  The observed giant arc near the brightest galaxies of the cluster is detected in the \emph{g}-image in the form of three separate sources. The arc lies nearly on one circle together with  two sources to the southeast (concentrically aligned with the arc in the HST image), what indicates a high degree of circular symmetry of the lens and its excellent position on the line of sight between the observer and the lensed galaxy.  Thus, to estimate  the lens mass we can consider a concentrically symmetric 2D mass distribution of the lens. Fig.~\ref{fig:SL_griz} shows this region in the \emph{g,r,i,z} bands. The blue and red circles show the fits of Einstein ring position obtained, respectively, by analyzing the HST image of the arc and the
southeastern arclets and the identified blue sources based on RTT-150 data.   The radius of the circle determines the angular distance  $\Theta_{arc}$ corresponding to the Einstein radius. The centers of the circles are also marked: blue - according to HST data, red - according to RTT150 data. It can be seen that the radii of the circles and their centers are in good agreement with each other. Thus, we demonstrated the possibility of using RTT-150 in
estimates of this kind. We took the distance between the center of the circle and \emph{BCG}, $1\arcsec.3$, as the error of the $\Theta_{arc}$ determination in this system.

\section{The redshift estimation of the lensed source}

\begin{figure}
  \centering

   \includegraphics[width=\columnwidth]{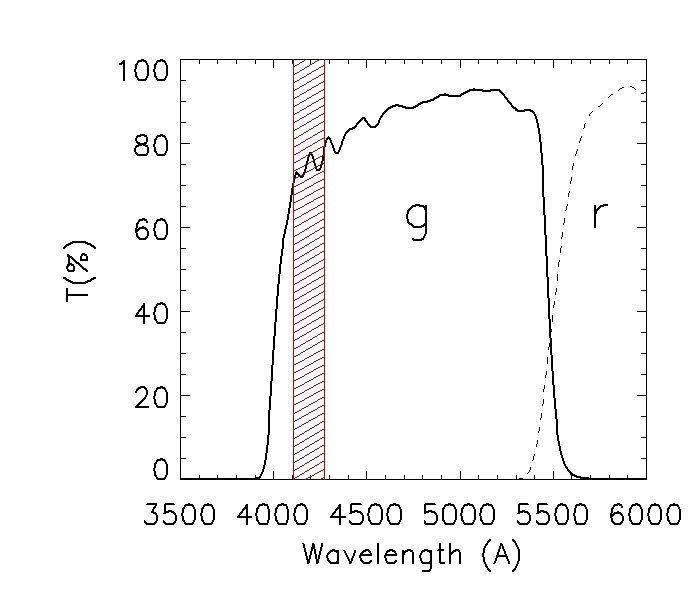}

  \caption{Transmission curve of the \emph{g} filter used at  \mbox{RTT-150}. The wavelength region of the line position $L_{\alpha}$ of the distant source, at which the lines $L_{\alpha}$ and CIV simultaneously fall in the region of this band, is marked.}
  
  \label{fig:filter}
  
\end{figure}

The total exposure times for the deep fields in the \emph{g} and \emph{r} bands are the same. Consequently, taking into account the same quantum efficiency of the detector and filter transmission in these bands, the depth of the fields is also approximately the same. The absence of signal in the \emph{r,i,z} bands allowed us to constrain the redshift of the lensed source. With a great probability, it is a galaxy with intense star formation. Therefore, the signal registered in the \emph{g} band corresponds to the location of the emission line $L_{\alpha}$ (1215\AA) in this region of the spectrum. Moreover, the strong CIV line (1549\AA) must also fall into this region. Otherwise, the signal from it would be registered in the \emph{r} band. Also, the CIV cannot be the only line falling into the \emph{g} band. In this case, the strong MgII emission line (2799\AA) would be registered in the \emph{i} band. Based on these considerations and analyzing the transmission function of the \emph{g} filter used on RTT-150, we determined the lower and upper limits on the redshift of the lensed galaxy. In Fig.~\ref{fig:filter}, the red dashed line shows the wavelength region of the $L_{\alpha}$ distant source line position, at which the $L_{\alpha}$ and CIV lines simultaneously fall within the region of this band. When estimating the wavelength interval, the filter transmittance region above 70\% was considered.

Thus, the redshift of the lensed source was determined: 
\mbox{$z_{s} = 2.44 \pm 0.07$}, which was used in the mass estimate. 

\section{Mass estimation from strong gravitational lensing}

In this work, we use the same approach to mass estimation as in the work by \cite{2016A&A...590L...4D}, performed for the PSZ1 G311.65-18.48 cluster, which was discovered by the \textit{Planck} observatory. In the field of this cluster a gravitationally lensed image of a distant galaxy in the form of a giant arc was detected.
The parameters of the lensed system (Einstein's radius in angular $ \Theta_{arc} $ and physical units $ R_{Ein} $, redshifts to the source and the lens and the corresponding angular diameter distances\footnote{The angular diameter distance is defined as the ratio of the transverse size of an object in physical units to its angular size in radians.}) for calculating the \cl2305 mass are given in Table~\ref{tab:tbl_par}.

The mass of a cluster inside a cylinder of radius $ R_{Ein} = 72.2 $ kpc is estimated as:
\begin{equation}
\label{eq:Ein}
M^{cyl}_{Ein}= \pi R_{Ein}^2 \Sigma_{crit} = 3.3^{+0.95}_{-0.8} \cdot10^{13} M_\odot,
\end{equation}
where $ \Sigma_{crit} $ is the critical surface density (equation~(\ref{eq:sigma_crit}) in the Appendix).

The mass of the cluster  estimated in \cite{br21_pazh} is  $ M_{500} \sim 9.03 \cdot10 ^ {14} M_\odot $, which means that the radius, 
within which the  mean density is equal to 500 times the critical density
at $ z = 0.7573$, is equal to $ R^{X}_{500} \sim 1110\,$ kpc. That corresponds to the angular size of $ \sim 2.5 '$. Thus, to recalculate the strong lensing mass within $R_{Ein}$ to the value of $ M(<R^{X}_{500}) $, we need an extrapolation in radius by the factor of $ R^{X}_{500}/R_{Ein} \sim 15 $, what limits the accuracy of such calculations.

A rough estimate of the cluster mass  enclosed within a sphere of radius $ R^{X}_{500} $ can be obtained by assuming that the lensing cluster is described by a singular isothermal sphere (SIS) model with the density distribution $ \rho(r) \propto r^{- \gamma } $, where $ \gamma = 2 $. Similar calculations were performed for elliptical galaxies by \cite{2018MNRAS.475.2403L}. For a given $ \gamma $, the mass~(\ref{eq:Ein}) within a cylinder of radius $ R_{Ein} $ can be converted to the mass within a sphere of the same radius. For $ \gamma = 2 $, this mass is $ 2 \cdot 10^{13} M_\odot $. In that case, the mass of the cluster inside the sphere of radius  $1110$ kpc is
$ M = (3.3 \pm 0.4) \cdot 10^{14} M_\odot $, which is $ \sim 3 $ times less than the mass derived from the X-ray luminosity-mass relation in  \cite{br21_pazh}.

\begin{table}
\caption{ Parameters of the gravitationally lensed system: Einstein radius in angular ($\Theta_{arc}$) and physical ($R_{Ein}$) units, redshifts of the source $ z_s $ and the lens $ z_d $, and the corresponding angular diameter distances $ D_s $ and $ D_d $ } 
  \label{tab:tbl_par}
  \vskip 2mm
  \renewcommand{\arraystretch}{1.05}
  \renewcommand{\tabcolsep}{0.35cm}
  \footnotesize
 
     \begin{tabular}{lll}
     \noalign{\vskip 3pt\hrule\vskip 5pt}

    $\Theta_{arc}$	&   $9\arcsec.8 \pm 1\arcsec.3$	\\
    $R_{Ein}$		    &	$72.2 \pm 9.6$ kpc	\\ 
    $z_{s}$		    &	$2.44 \pm 0.07$    	\\
    $z_{d}$		    & 	$0.7573 \pm 0.0006$	  		\\
    $D_{s}$		    &	$1674_{+9.6}^{-9.9}$ Mpc   \\
    $D_{d}$		    &	$1519 \pm 0.44$ Mpc 	\\

\noalign{\vskip 3pt\hrule\vskip 3pt}
\end{tabular}
\end{table}

There is no doubt that the assumption about the slope $ \gamma = 2 $ over the entire range of radii from $ \sim 0.07 $ $ R^X_{500} $ to $ R^X_{500} $ is a rough approximation of the actually observed density profiles. It can be used only for an order of magnitude estimation. The Navarro-Frank-White profile \citep{1996ApJ...462..563N} or the Einasto profile \citep{1965TrAlm...5...87E}, which describe well the numerical simulations of cluster formation when the contribution of ordinary matter (baryons) can be neglected, are a better approximation. In the case of a slightly simpler Navarro-Frank-White profile (compared to the Einasto model), the slope of the density profile changes from $ \gamma = 1 $ at small radii to $ \gamma = 3 $ at large radii.

Let us  estimate the mass of the cluster assuming that the density obeys the Navarro-Frenk-White model:
\begin{equation}
    \rho(r) = \frac{\delta_c \rho_{crit}}{\left(\frac{r}{R_s}\right)\left(1+\frac{r}{R_s}\right)^2},
    \label{eq:nfw}
\end{equation}
where $ \rho_{crit} = 3 H^2 (z) / (8 \pi G) $ is the critical density, $ H(z) = H_0(\Omega_m (1 + z)^3 + \Omega_{\Lambda})^{1/2} $ is the Hubble constant at the redshift of the cluster, $ G $ is the gravitational constant. The characteristic cluster radius is $ R_s = R_{200}/c $, where $c$ is the dimensionless concentration parameter, which is included in the normalization of the density profile as
\begin{equation}
\delta_c = \frac{200}{3} \frac{c^3}{ln(1+c) - c/(1+c)}.
\end{equation}
The size of a cluster  $ R_{200} $  is defined as the radius of a sphere within which the mean halo density is $ 200 \rho_{crit} $. Therefore, the halo mass is $M_{200} \equiv M(R_{200}) = \frac{4}{3} \pi R^3_{200} \times 200\rho_{crit}$.

\begin{figure}
  \centering
   \includegraphics[width=\columnwidth]{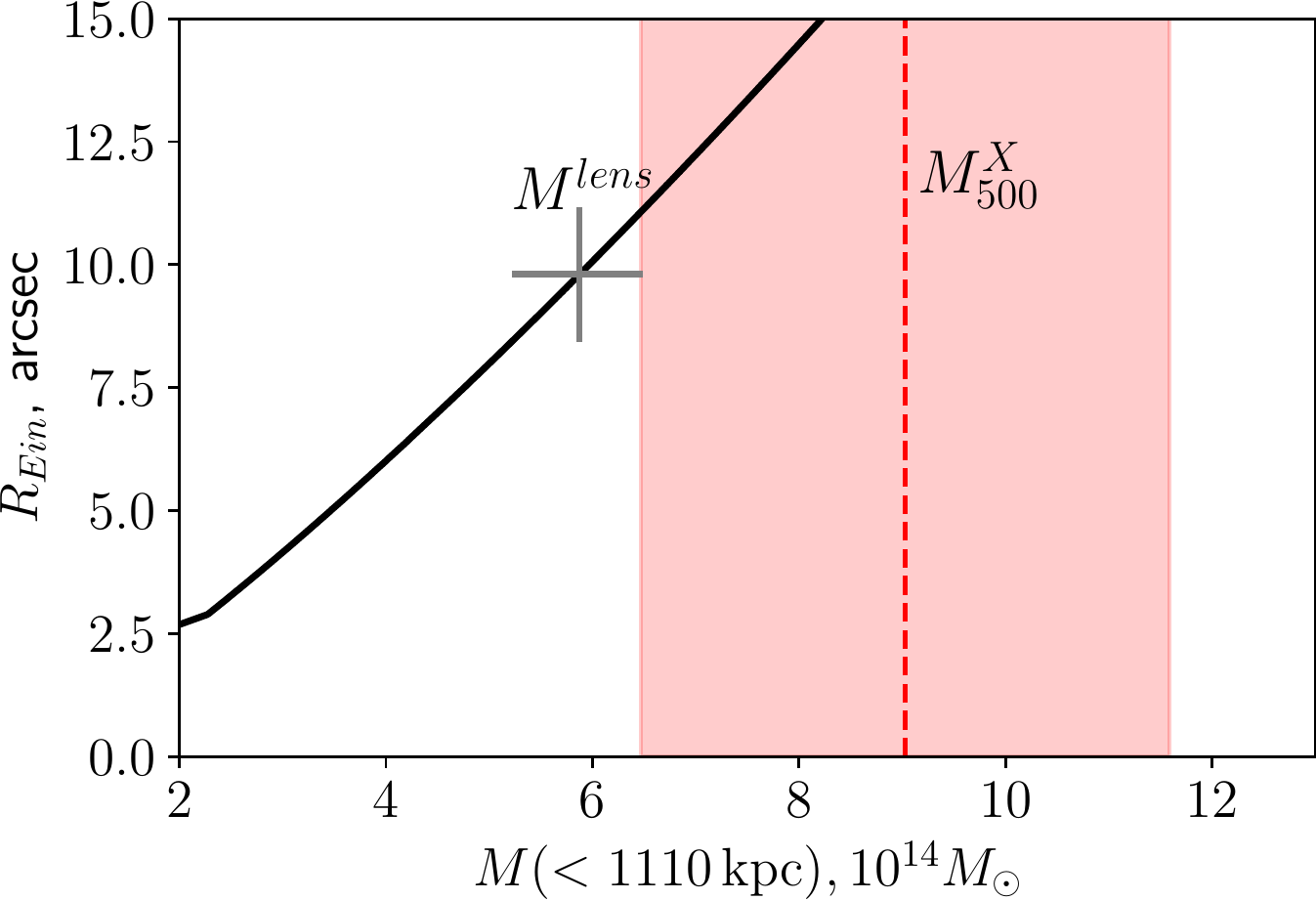}
  \caption{Dependence of the size of the tangential critical curve for the spherically symmetric Navarro-Frenk-White profile on the cluster mass estimated within the sphere of radius $ R_{500}^X = 1110 $ kpc. The concentration parameter $ c $ is related to the halo mass by the relation from  \cite{2008MNRAS.390L..64D} derived from the analysis of cosmological simulations without taking baryons into account. The gray cross denotes the mass of the cluster at which the Einstein radius coincides with the radius of the circle $ \Theta_{arc} $. When estimating the uncertainty for the mass, only the error in the determination of $ \Theta_{arc} $ was taken into account. The vertical dashed line and shaded area show the mass estimate for the SRGe cluster CL2305.2-2248 from SRG/eROSITA data.}
  \label{fig:nfw}
\end{figure}

Studies of the properties of dark matter halos based on numerical simulations show \citep[i.e.,,][]{2008MNRAS.390L..64D} that the parameters $c$ and $M_{200}$ are closely related to each other in a wide range of halo masses. For further estimates, we use the  halo concentration-mass relation (taking into account the redshift of the cluster) from  \cite{2008MNRAS.390L..64D}. Thus, we are left with  one free parameter in the Navarro-Frenk-White model - the halo mass. For a spherically symmetric lens, the position of the tangential arc should be close to the tangential critical curve \citep[the properties of the lens with the Navarro-Frenk-White density profile are described in][also see Appendix]{1996A&A...313..697B}. By equating $ \Theta_{arc} $ to the size of the critical curve in the lens plane, we obtain an independent estimate for the halo mass. Under the assumptions above, the best agreement with the observed position of the arc is achieved at $ M_{200} = 7.1 \cdot 10^{14} M_{\odot} $ ($ R_{200} = 1388 $ kpc). In this case, the concentration parameter is equal to $ c = 3.8 $ and $ R_s = R_ {200} / c = 365 $ kpc.
Thus, for comparison with the mass estimates from X-rays and the Sunyaev-Zeldovich effect, we can calculate the mass of the dark matter halo $ M_{500} = (4.9 \pm 0.7) \cdot 10^{14} M_{\odot} $ within the radius inside which the mean halo density is equal to  $ 500 \rho_ {crit} (z = 0.7573) $, as well as the mass within $ R^{X}_{500} $ = 1110 kpc, which is $ M (<R^{X}_{500}) = (5.9 \pm 0.6) \cdot 10^{14} M_{\odot} $. Here, the uncertainties in mass estimates are based on the uncertainty in the position of the halo center, i.e. the center of the circle (see Fig.~\ref{fig:SL_griz}). The radius of tangential critical curve for the Navarro-Frenk-White profile is plotted against the dark halo mass $ M(<R^{X}_{500}) $ in Fig.~\ref{fig:nfw}. The mass estimate that matches best  the position of the tangential arc is marked with a grey cross. For comparison, we also show the cluster  mass   within $ R^{X}_{500} = 1110 $~kpc obtained in \cite{br21_pazh} from X-ray observations, performed within the framework of the SRG/eROSITA survey. The mass estimates $ M_{500} $ of the SRGe CL2305.2-2248 cluster available to date are presented in Table~\ref{tab:masses}. Note that the physical size of the sphere inside which the mass is calculated can vary for different methods. As seen from Table~\ref{tab:masses} and Fig.~\ref{eq:nfw}, the mass of the cluster estimated from gravitational lensing is $ \sim 1.5-2 $ times smaller than the mass  from the literature.

Note that the size of the critical curve substantially depends on the derivative of the gravitational potential in the central part of the cluster. Consequently, the strong lensing mass estimate may turn out to be sensitive to the contribution of central galaxies and, possibly, hot X-ray emitting gas, to the total density profile of a cluster. Numerical cosmological simulations show that taking baryons into account can indeed significantly change the density profile in the central region of the cluster \citep[for example,][]{2017MNRAS.465.3361H, 2018MNRAS.477.2804S}, mainly due to cooling and "adiabatic compression" \citep[for example,][]{2004ApJ...616...16G}. These effects can be partially taken into account by keeping the NFW radial profile (\ref{eq:nfw}), but changing the value of the concentration parameter $c$ compared to expectations from cosmological dark matter only simulations. Above we obtained the cluster mass estimate  under the assumption that the concentration parameter is expressed via the halo mass according to the relation presented in \cite{2008MNRAS.390L..64D}. However, observations of strong and weak lensing of several well-studied clusters from the CLASH sample \citep{2015ApJ...806....4M} give noticeably larger values of $ c $ than predicted in \cite{2008MNRAS.390L..64D}, where  cosmological simulations involving only dark matter were used. If we fix the concentration parameter at the value that is two times higher than the predictions from \cite{2008MNRAS.390L..64D}, then the estimate of the halo mass is $ M_{200} = 2.6 \cdot 10^{14} M_{\odot} $ , $ c = 8.3 $, $ M_{500} = 2.1 \cdot 10^{14} M_{\odot} $, $ M(<R^X_{500}) = 2.8 \cdot 10^{14} M_{\odot} $. Thus, there is a noticeable uncertainty when extrapolating from $ R_{Ein} = 72 $ kpc to large radii $ \sim1000 $ kpc due to the fact that real profiles of the total density of galaxy clusters can be much more complicated than predicted by numerical simulations of cluster formation without baryons, and due to possible deviations from spherical symmetry.

\begin{table}
\caption{SRGe CL2305.2-2248 cluster mass estimates from the X-ray and microwave observations and   strong lensing} 
  \label{tab:masses}
  \vskip 2mm
  \renewcommand{\arraystretch}{1.05}
  \renewcommand{\tabcolsep}{0.35cm}
  \footnotesize
 
     \begin{tabular}{cclcc}
     \noalign{\vskip 3pt\hrule\vskip 5pt}

     & 	$M_{500}$, 10$^{14}M_{\odot}$	&	Source	 \\
\hline
\noalign{\vskip 3pt}

eROSITA	&   $9.0\pm 2.6$	&   \cite{br21_pazh} \\
ACT	&   $9.2 \pm 1.5 $	&   \cite{2021ApJS..253....3H} \\
SPT	&  $7.4\pm 0.8$	&   \cite{2020ApJS..247...25B}	 \\
strong 	&   $4.9\pm 0.7$	&   this work \\
lensing	&   	&   \\
\noalign{\vskip 3pt\hrule\vskip 3pt}

\end{tabular}
\end{table}

\section{Discussion}

The massive galaxy cluster SRGe CL2305.2-2248 at redshift $ z \simeq 0.76 $ is a very rare object in the observable universe. Its mass, obtained from X-ray observations and from the Sunyaev-Zeldovich effect, $ M_{500} \sim 9 \cdot 10^{14} M_{\odot} $ is comparable to the mass of the El Gordo cluster. Within the framework of the standard cosmological model $\Lambda$CDM, only $ \sim 10 $ of such massive clusters at redshifts $ z> 0.7 $ are expected in the entire sky.
In this work, we investigated deep direct images of the field of the cluster \cl2305 \ obtained with the Russian-Turkish 1.5-m telescope (RTT-150) as a part of the tasks of ground support for the X-ray survey of the eROSITA telescope onboard the Spectrum-Roentgen-Gamma (SRG) space observatory. From the RTT-150 images and high spatial resolution images from the Hubble Space Telescope (HST) Open Data Archive, probable gravitationally lensed sources were identified, including the giant arc near the cluster brightest galaxies. The photometric redshift of the arc is estimated to be $ z_S = 2.44 \pm 0.07 $.
The arc and several other lensed images lie virtually on one circle with a radius of $ 9.8 \pm 1.3 $ arc seconds, which made it possible to make an assumption about the spherical symmetry of the lens and obtain a direct measurement of the mass inside a cylinder with a radius equal to the radius of the circle (= Einstein's radius $ R_{Ein} $).
Extrapolation of mass measurements from strong lensing, that uses the singular isothermal sphere or the Navarro-Frenk-White profile approximation for a reasonable range of values of the concentration parameter $ c $, leads to estimates of $ M_{500} $ approximately $1.5-3$ times lower than the estimate from the relation $ M-L_X $. Values of $c$ that
are well below the typical ones for relaxed clusters
are required to reconcile these estimates. A possible
explanation is a merger of clusters that led to a flatter
mass distribution in the central part of the cluster.
  The presence of two massive galaxies in the center may also argue for this hypothesis. Another explanation could be a peculiarly high X-ray luminosity of this cluster. In principle, the luminosity can increase noticeably during certain merging phases. Consequently, the merger process can simultaneously lead to an overestimation of the luminosity and an underestimation of the concentration parameter $c$. Detailed  X-ray and microwave observations are needed to reliably determine the dynamic state of the cluster. Constructing an accurate lens model using strong and weak lensing methods will also allow one to draw conclusions about whether we  observe one massive cluster or several mergers, and to estimate the halo mass(es) regardless of the dynamic state. However, this requires measuring/evaluating redshifts of a large number of sources and measuring the degree of elongation of background objects (for weak lensing).

\acknowledgements

This work is based on observations with the eROSITA telescope on board the SRG observatory. The SRG observatory was built by Roskosmos in the interests of the Russian Academy of Sciences represented by its Space Research Institute (IKI) within the framework of the Russian Federal Space Program, with the participation of the Deutsches Zentrum f{\''u}r Luft- und Raumfahrt (\emph{DLR}). The SRG/eROSITA X-Ray telescope was built by a consortium of German Institutes led by the Institute of Extraterrestrial Physics (\emph{MPE}) with the support from \emph{DLR}. The SRG spacecraft was designed, built, launched and is operated by the Lavochkin Association and his subcontractors.
The science data are downlinked via the Deep Space Network Antennae in Medvezhye Lakes, Ussuriysk, and Baikonur, funded by Roskosmos. The eROSITA data used in this work were processed using the \emph{eSASS} software system developed by the German eROSITA consortium and proprietary data reduction and analysis software developed by the Russian eROSITA Consortium. The authors thank TUBITAK, IKI, KFU and AN RT for support of observations on the Russian-Turkish 1.5-m telescope in Antalya (RTT-150).

The study is based on observations made with the NASA/ESA Hubble Space Telescope and obtained from the Hubble Legacy Archive, which is a joint project of the Space Telescope Science Institute (STScI/NASA), the Space Telescope European Coordinating Facility (ST-ECF/ESAC/ESA) and the Canadian Astronomy Data Center (CADC/NRC/CSA).

The work of IFB, RAB, SND was supported by the Russian Science Foundation, grant 21-12-00210. AAS was partially supported by the project 0033-2019-0005 of the Ministry of Education and Science of the Russian Federation.

\bibliographystyle{astl}
\bibliography{eng_cl23052m2248_SL}

\section*{Appendix}

Analytical expressions which describe the position of tangential arcs in a spherically symmetric gravitational lensing system, in which the lens density profile is described by the Navarro-Frenk-White model, are presented in \cite{1996A&A...313..697B}. The properties of gravitational lenses with axial symmetry are determined by their surface density $ \kappa (x) = \Sigma(x) / \Sigma_{cr} $, where
\begin{equation}\label{eq:sigma_crit}
\Sigma_{\mathrm{cr}}=\frac{c^{2}}{4 \pi G} \frac{D_{\mathrm{s}}}{D_{\mathrm{d}} D_{\mathrm{ds}}}, 
\end{equation}
$ x $ is the coordinate along the radius in $ R_s $ units, $ c $ is the speed of light, $ G $ is the gravitational constant, $ D_d $, $ D_s $ is the angular diameter distance from the observer to the lens and to the source, respectively, $ D_{ds} $ - angular diameter distance from the lens to the source. The mass enclosed within a circle of radius $ x $ is given by
\begin{equation}
m(x) \equiv 2 \int_{0}^{x} d y y \kappa(y).
\end{equation}
In case of spherically symmetric lenses, the source images in the form of tangential arcs appear close to the tangential critical curve, which is determined by the condition $ m(x) = x^2 $. The volumetric density $ \rho(r) $, defined by the formula~(\ref{eq:nfw}), corresponds to the dimensionless surface density
\begin{equation}
\kappa(x)=2 \kappa_{\mathrm{s}} \frac{f(x)}{x^{2}-1},
\end{equation}
where $\kappa_{\mathrm{s}} \equiv \rho_{\mathrm{s}} r_{\mathrm{s}} \Sigma_{\mathrm{cr}}^{-1}$. 
The dimensionless mass $ m(x) $ is given by the expression
\begin{equation}
m(x)=4 \kappa_{\mathrm{s}} g(x),
\end{equation}
where $$g(x)=\ln \frac{x}{2}+\left\{\begin{array}{ll}\frac{2}{\sqrt{x^{2}-1}} \arctan \sqrt{\frac{x-1}{x+1}} & (x>1) \\ \frac{2}{\sqrt{1-x^{2}}} \mathrm{arctanh} \sqrt{\frac{1-x}{1+x}} & (x<1) \\ 1 & (x=1)\end{array}\right.$$

\end{document}